\documentstyle[12pt,epsfig]{article}
\begin{document}

\newcommand{\lsim}{\raisebox{-0.13cm}{~\shortstack{$<$ \\[-0.07cm] $\sim$}}~}
\newcommand{\gsim}{\raisebox{-0.13cm}{~\shortstack{$>$ \\[-0.07cm] $\sim$}}~}
\newcommand{\nn}{\noindent}
\newcommand{\non}{\nonumber}
\newcommand{\tb}{\tan \beta}
\newcommand{\s}{\smallskip}
\newcommand{\beq}{\begin{eqnarray}}
\newcommand{\eeq}{\end{eqnarray}}
\newcommand{\etal}{{\it et al.}\ }

\vspace*{.1cm} 
\baselineskip=15.7pt

\begin{flushright}
ETHZ--IPP PR--2003--01\\
CERN TH/2003--145\\
PM/03--14\\
27 June 2003\\
\end{flushright}

\vspace*{0.9cm}

\begin{center}

{\large\sc {\bf Z$^\prime$ studies at the LHC: an update}}

\vspace{0.7cm}

{\sc Michael DITTMAR}, {\sc Anne-Sylvie NICOLLERAT}\s

Institute for Particle Physics IPP, ETH Z\"urich, CH-8093 Z\"urich,
Switzerland. \bigskip

and \bigskip

{\sc Abdelhak DJOUADI}\s

Theory  Division, CERN, CH-1211 Geneva 23, Switzerland.\s

Laboratoire de Physique Math\'ematique et Th\'eorique, UMR5825-CNRS,\\
Universit\'e de Montpellier II, F-34095 Montpellier Cedex 5, France. 
\end{center} 

\vspace*{1cm} 

\begin{abstract}

\nn We reanalyse the potential of the LHC to discover new $Z'$ gauge bosons and
to discriminate between various theoretical models. Using a fast LHC detector
simulation, we have investigated how well the  characteristics of $Z'$ bosons
from different models can be measured. For this analysis we have combined  the
information coming from the cross section measurement, which provides also the
$Z'$ mass and total width, the forward-backward charge asymmetries on- and
off-peak,  and the $Z'$ rapidity distribution, which is sensitive to its $u
\bar{u}$ and $d \bar{d}$ couplings. We confirm that new $Z'$ bosons can be
observed in the process $pp \to Z' \to \ell^+ \ell^-$, up to masses of about 5
TeV for  an integrated luminosity of 100 fb$^{-1}$. The off- and on-resonance
peak forward-backward charge asymmetries  $A_{\rm FB}^{\ell}$  show that
interesting statistical accuracies can be obtained up to  $Z'$ masses of the
order of 2 TeV. We then show how the different experimental observables  allow
for a diagnosis of the $Z'$ boson and the distinction between the various
considered models.

\end{abstract}
\vspace*{1.8cm}

\newpage 

\subsection{Introduction}

Although the Standard Model (SM) of the electroweak and strong interactions 
describes nearly all experimental data available today \cite{High-precision},
it is widely believed that it is not the ultimate theory. Grand Unified
Theories (GUTs), eventually supplemented by Supersymmetry to achieve a
successful unification of the three gauge coupling constants at the high scale,
are prime candidates for the physics beyond the SM. Many of these GUTs,
including superstring and left-right-symmetric models, predict the existence of
new neutral gauge bosons, which might be light enough to be accessible at
current and/or future colliders\footnote{For example the breaking at the
supersymmetry-breaking scale, i.e. at a scale around the TeV, of an extra 
U(1)$\prime$ group to which a $Z'$ boson is associated, might solve the
so-called $\mu$ problem, which  notoriously appears in the Minimal
Supersymmetric extension of the SM  \cite{Cvetic}.}; for reviews see
Ref.~\cite{Reviews}.  New vector bosons also appear in models of dynamical
symmetry breaking \cite{dynamical} and recently, ``little Higgs" models have
been proposed to solve the hierarchy problem of the SM \cite{LittleH}: they
have large gauge group structures and therefore predict a plethora of new gauge
bosons with masses in the TeV range. \s 

The search for these $Z'$ particles is an important  aspect of the experimental
physics program of future  high-energy colliders. Present limits from direct
production at the Tevatron and  virtual effects at LEP, through interference or
mixing with the $Z$ boson, imply that new $Z'$ bosons are rather heavy and mix
very little with the $Z$ boson.  Depending on the considered theoretical 
models, $Z'$ masses of the order of 500 to 800 GeV and $Z$--$Z'$ mixing angles
at the level of a few per-mile are  excluded\footnote{In contrast, some
experimental data on atomic parity violation and deep inelastic
neutrino-nucleon scattering, although controversial and of small statistical
significance [see Ref.~\cite{High-precision} for instance], can be explained
by  the presence of a $Z'$ boson \cite{Anomaly}.}  \cite{Langacker}.  A $Z'$
boson, if lighter than about 1 TeV, could be discovered at Run II  of the
Tevatron \cite{Theory} in the Drell-Yan process  $p\bar{p} \to Z' \to \ell^+
\ell^-$, with $\ell=e,\mu$ \cite{DY}.  Detailed theoretical \cite{Theory} and
experimental \cite{aachenlhc,Exp1,Exp2} analyses have shown that the discovery
potential of the LHC experiments is about  5 TeV, using the process $pp \to Z'
\to \ell^+ \ell^-$. Future $e^+e^-$ colliders with high c.m.\ energies and
longitudinally polarized beams could indicate the existence of $Z'$ bosons via
its interference effects, with masses up to about $6\times \sqrt{s}$
\cite{Theory,eepaper}.\s

After the discovery of a $Z'$ boson, some  diagnosis of  its coupling needs to
be done  in order to identify the correct theoretical frame. For this purpose,
and since a long time, the forward-backward charge asymmetry  for leptons
$A_{\rm FB}^{\ell}$ has been advocated as being a powerful tool  \cite{AFB};
the most direct method to actually measure $A_{\rm FB}^{\ell}$ at the LHC has
been described in~\cite{Previous}. In addition to the information from the
total $Z'$ cross section,  it has been argued that the measurement of ratios of
$Z'$ cross sections in different rapidity bins might provide some information
about the $Z'$ couplings to up and down quarks \cite{rapidity}.\s

Following the arguments given in \cite{lhclumi}, we advocate that the  $Z'$
cross section should be measured relative to the number of produced  $Z$ bosons
for the same lepton final states.  Using this approach, many systematic
uncertainties due to theoretical and experimental uncertainties will cancel,
and  the relative $Z'/Z$ cross section ratio might be measured and calculated
with an accuracy of about 1\%.  Furthermore, the method  should also lead to
precise  relative parton distribution functions for $u$ and $d$ quarks, as well
as for the corresponding  sea quarks and antiquarks.  Thus, we can go beyond
the previously proposed procedure  to analyse the $Z'$ rapidity distribution
\cite{rapidity}, by performing a fit. The fit uses the predicted rapidity
spectra as calculated with $u\bar{u}$ and $d\bar{d}$, as well as the
contribution of the sea, for the mass region of interest, which is directly 
related to $x_{1}, x_{2}$ of the  corresponding quarks and antiquarks in
the proton.\s

While numerous theoretical and experimentally motivated $Z'$ studies  have
already  been performed, the combination of all sensitive LHC  variables, as
described above, has not been done so far; the work described in this paper
will thus fill a gap.  We will perform the studies using the {\tt PYTHIA}
program \cite{PYTHIA} and a fast LHC detector simulation. We first update
previous studies using, the latest parton distribution functions \cite{cteq}, 
and extend them in two directions. First, following the method proposed 
in~\cite{Previous}, the forward-backward charge asymmetries,  on and off the
$Z'$ resonance peak, are analysed together with the cross section in order to
differentiate between the different models\footnote{Recently, the off-peak
forward-backward asymmetry has also been used in Ref.~\cite{Georges} to study
Kaluza-Klein excitations of gauge bosons.}. Second, we show that a direct fit
of the rapidity distribution allows for additional information and  would be
useful to disentangle between $Z'$ bosons from various models through their
different couplings to up-type and down-type quarks. \s

The rest of the discussion will be organized as follows. In the next section,
we define the theoretical framework in which our analysis will be performed. In
section 3, we describe the relevant observables that can be measured at the
LHC, namely the dilepton cross section times the $Z'$ total width, the on-peak
and off-peak  forward-backward asymmetries and the rapidity distribution, 
and the simulation tools which we will use in our study. In section 4, we
analyse the resolving power of these observables.

\subsection{The considered $Z'$ models} 
\label{models}
To simplify the discussion, we will focus in this paper on two effective
theories of well motivated models that lead to an extra gauge boson: \s

1) An effective ${\rm SU(2)_L \times U(1)_Y \times U(1)_{Y'}}$ model, which
originates from  the breaking of the exceptional group E$_6$, which is general
enough to include many interesting possibilities. Indeed, in the breaking of
this group down to the SM symmetry, two additional neutral gauge bosons could
appear. For simplicity we assume that only the lightest $Z'$ can be produced at
the LHC. It is defined as
\beq
Z' = Z'_\chi \cos\beta + Z'_\psi \sin \beta
\eeq
and can be parametrized in terms of the hypercharges of the two groups
U(1)$_\psi$  and U(1)$_\chi$ which are involved in the breaking chain: 

$ \hspace*{-0.5cm}
{\rm E_6 \to SO(10) \times U(1)_\psi \to SU(5) \times U(1)_\chi \times U(1)_\psi
\to SU(3)_c \times SU(2)_L \times U(1)_Y \times U(1)_{Y'} } 
$

The values $\beta=0$ and $\beta=\pi/2$ would correspond, respectively, to pure
$Z'_\chi$ and $Z'_\psi$ bosons, while the value $\beta= {\rm
arctan}(-\sqrt{5/3})$ would correspond to a $Z'_\eta$ boson originating from
the direct breaking of ${\rm E_6}$ to a rank-5 group in superstrings inspired
models.\s

2) Left-right (LR) models, based on the symmetry group  ${\rm SU(2)_R \times
SU(2)_L \times U(1)_{B-L}}$, where $B$ and $L$ are the baryon and lepton
numbers. Even though we investigate only the $Z'$ in this paper, it should be
recalled that new charged vector bosons, potentially observable at the LHC, also
appear in these models. The most general neutral boson $Z'_{LR}$ will couple to
a linear  combination of the right-handed and $B$--$L$ currents:
\beq
 J_{LR}^\mu = \alpha_{LR} J^\mu_{3R} - (1/2\alpha_{LR}) J^\mu_{B-L}
\ \ {\rm with} \ \alpha_{LR}= \sqrt{ (c_W^2 g_R^2/s_W^2 g_L^2)-1} \ , 
\eeq
where $g_L$=$e/s_W$ and $g_R$ are the ${\rm SU(2)_L}$ and ${\rm SU(2)_R}$
coupling constants with $s_W^2= 1-c_W^2 \equiv \sin^2\theta_W$. The parameter
$\alpha_{LR}$ is restricted to lie in the range $\sqrt{2/3} \lsim \alpha_{LR}
\lsim \sqrt{2}$: the upper bound corresponds to a LR-symmetric model with
$g_R=g_L$, abbreviated in the following as $LR$, while the lower bound
corresponds to the $\chi$ model discussed in  scenario (1), since SO(10) can 
lead to both ${\rm SU(5) \times U(1)}$ and ${\rm SU(2)_R \times SU(2)_L \times
U(1)}$ breaking patterns. \s

In order to achieve a complete comparison, we will also discuss the
non-realistic case of a sequential boson $Z'_{\rm SM}$, which has the same fermion
couplings as the SM $Z$ boson, as well as a $Z'$ boson, denoted by  $Z'_{d}$, 
with vanishing axial and vectorial couplings to $u$ quarks and which, in  $E_6$
models, corresponds to the choice $\cos\beta=\sqrt{5/8}$. \s

The left- and right-handed couplings of the $Z'$ boson to fermions, defined
as: 
\beq
g_{Z'} J^\mu_{Z'} Z'_\mu = \frac{e}{c_W} \sum_f \gamma^\mu \left[ 
\frac{1-\gamma_5}{2} \, g_L^{fZ'} + \frac{1+ \gamma_5}{2} \, g_R^{fZ'} \right]
\ , 
\eeq
are given in Table 1 for the first-generation fermions in the two scenarios. 
The mixing between the $Z$ and $Z'$ bosons is very small 
\cite{Langacker} and will be neglected in our discussion.

\begin{table}[htbp]
\vspace*{-5mm}
\begin{center}
\renewcommand{\arraystretch}{1.7}
\begin{tabular}{|c||c|c||c|c|}
\hline
$f$  & $g_L^{fZ'}|_{E_6}$ & $g_R^{fZ'}|_{E_6}$ & $g_L^{fZ'}|_{LR}$ & 
$g_R^{fZ'}|_{LR} $ \\
\hline \hline
$\nu_e$ & $\frac{3 \cos\beta}{2\sqrt{6}}+ \frac{ \sqrt{10}\sin\beta}{12}$
& $0$ & $\frac{1}{2\alpha_{LR}}$ & $0$ \\ \hline
$e$ & $\frac{3 \cos\beta}{2\sqrt{6}}+ \frac{ \sqrt{10}\sin\beta}{12}$
& $\frac{\cos\beta}{2\sqrt{6}}- \frac{ \sqrt{10}\sin\beta}{12}$
& $\frac{1}{2\alpha_{LR}}$ & $\frac{1}{2\alpha_{LR}}- \frac{\alpha_{LR}}{2}$ 
\\ \hline
$u$ & $-\frac{\cos\beta}{2\sqrt{6}}+ \frac{ \sqrt{10}\sin\beta}{12}$
& $\frac{\cos\beta}{2\sqrt{6}}- \frac{ \sqrt{10}\sin\beta}{12}$
& $-\frac{1}{6\alpha_{LR}}$ & $-\frac{1}{6\alpha_{LR}}+ \frac{\alpha_{LR}}{2}$ 
\\ \hline
$d$ & $-\frac{ \cos\beta}{2\sqrt{6}}+ \frac{ \sqrt{10}\sin\beta}{12}$
& $-\frac{3\cos\beta}{2\sqrt{6}}- \frac{ \sqrt{10}\sin\beta}{12}$
& $-\frac{1}{6\alpha_{LR}}$ & $-\frac{1}{6\alpha_{LR}}- \frac{\alpha_{LR}}{2}$ 
\\ \hline
\end{tabular}
\end{center}
\vspace*{-3mm}
\caption[]{\it Left- and right-handed couplings of the $Z'$ boson to the
SM fermions with the notation of the first generation in the $E_6$ (left panels) 
and LR (right panels) models.}
\vspace*{-3mm}
\end{table}

The $Z'$ partial decay width into a massless fermion-antifermion pair reads
\beq \Gamma_{Z'}^f = N_c \frac{\alpha M_{Z'}}{6 c_W^2} \left[ (g_L^{fZ'})^2 + 
(g_R^{fZ'})^2 \right] \eeq with $N_c$ the colour factor and the electromagnetic
coupling constant to be  evaluated at the scale $M_{Z'}$ leading to $\alpha \sim
1/128$. In the absence of any exotic decay channel, the branching fractions for
decays into the first-generation leptons and quarks are shown in Fig.~1 for
$E_6$ and LR models as functions of $\cos\beta$ and $\alpha_{LR}$,
respectively.   As can be seen, the decay fractions into $\ell^+ \ell^-$ pairs
are rather  small, varying between 6.6\% and 3.4\% for $E_6$ models and 6.6\%
and 2.3\% for LR models; in the latter case the decay branching fraction is
largest for the symmetric case $g_L=g_R$ and smallest for $\alpha_{LR}\simeq
\sqrt{2}$. The $Z'$ total decay width, normalized to $M_{Z'}$, is also shown in
Fig.~1: it is largest when $\cos\beta = \pm 1$  in $E_6$ models and
$\alpha_{LR} \simeq  \sqrt{2}$ in LR ones. The $Z'$ bosons that we will
consider here are thus  narrow resonances, since their  total decay width does
not exceed 2\% of their masses\footnote{Note however that  non-standard decays,
such as decays into supersymmetric  particles and/or decays into exotic
fermions, are possible;  if kinematically  allowed, they can increase the 
total decay width and hence decrease the $Z' \to \ell^+ \ell^-$  branching 
ratios. In the case of the $E_6$ model for instance, the fermions
belong to a representation of dimension {\bf 27} which contains 12 new
heavy states per generation, and if they are light enough, the total decay 
width of the $Z'$ is then simply  $\Gamma_{Z'} \simeq 2.5 M_{Z'}$\% 
independently of the angle $\beta$ \cite{AFB}. These exotic fermions, 
however, should also be observed at future colliders; see e.g. \cite{exotic}.}.

\begin{figure}[htbp]
\vspace*{-1.4cm}
\begin{center}
\hspace*{-1cm}
\mbox{\psfig{figure=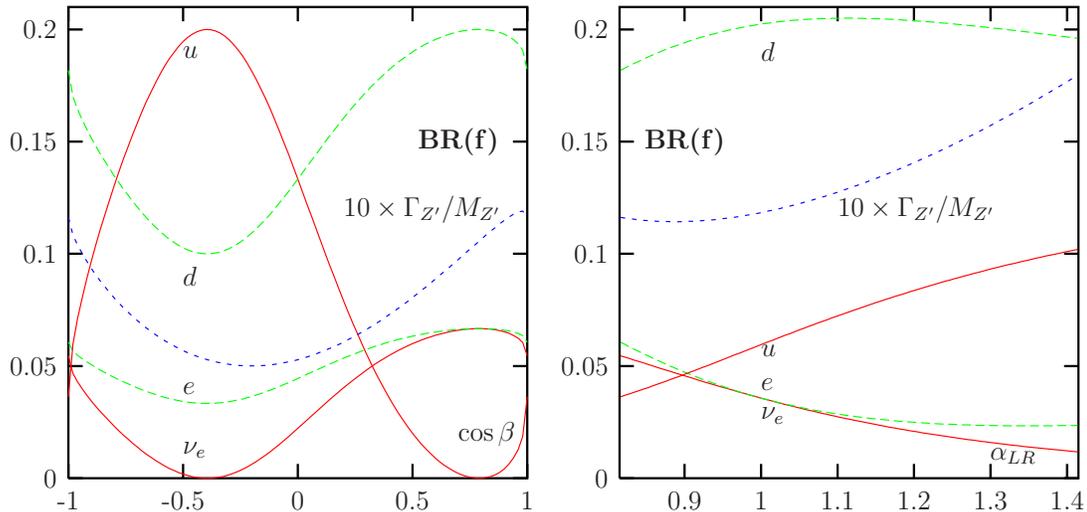,width=18cm}}
\end{center}
\vspace*{-18.1cm}
\caption[]{\it The branching ratios of the decays $Z' \to f\bar{f}$ in $E_6$ 
models as a function of $\cos\beta$ (left) and in LR models as a function of 
$\alpha_{LR}$ (right). The total $Z'$ decay widths, normalized to $10/M_{Z'}$, 
are also shown.}
\vspace*{-3mm}
\end{figure}

In the limit of negligible fermion masses, the differential cross section for 
the subprocess $q \bar{q} \to \ell^+ \ell^-$, with respect to $\theta^*$
defined as the angle between the initial quark $q$ and the final lepton
$\ell^-$ in the $Z'$ rest frame, is given by $[\hat{s}=M^2_{\ell \ell}$ is the
c.m.\, energy of the  subprocess]
\beq
\frac{ {\rm d}\hat{\sigma}}{ {\rm d}\cos \theta^*} \, (q\bar{q} \to \gamma, Z, 
Z'\to \ell^+ \ell^-) = \frac{1}{9} \frac{\pi \alpha^2}{2\hat{s}} \left[ (1+ 
\cos^2\theta^*) Q_1 +  2 \cos \theta^* Q_3 \right] \ ,
\eeq
where the charges $Q_1$ and $Q_3$ are given by \cite{eepaper}
\beq
Q_{1/3}= \left[ \, |Q_{LL}|^2 +| Q_{RR}|^2 \pm |Q_{RL}|^2 \pm |Q_{LR}|^2 \, 
\right] /4 \ .
\eeq
In terms of the left- and right-handed couplings of the $Z'$ boson defined
previously, and of those of the $Z$ boson [$g_L^{fZ}=I_{3L}^f-Q^f s_W^2, \,
g_R^ {fZ}=-Q_f s_W^2$] and the photon [$g_L^{f\gamma}=g_R^{f\gamma} = Q_f$]
with  $Q^f$ the electric charge and $I_{3L}^f$ the left-handed weak isospin,
the helicity  amplitudes $Q_{ij}$ with $i,j=L,R$ for a given initial $q\bar{q}$
state read
\beq
Q_{ij}^q= g_i^{q\gamma} g_j^{\ell \gamma} + \frac{g_i^{qZ} g_j^{\ell Z} }{
s_W^2 c_W^2}  \frac{\hat{s}} {\hat{s}- M_Z^2 + i\Gamma_Z M_Z} +
\frac{g_i^{qZ'} g_j^{\ell Z'} }{c_W^2}  \frac{\hat{s}} {\hat{s} - M_{Z'}^2 
+ i\Gamma_{Z'} M_{Z'} } \ .
\eeq
To obtain the total hadronic cross section\footnote{A $K$-factor of the order
of $K_{\rm DY} \sim 1.4$ \cite{Altarelli} for the production cross section  can
be  also included.} and forward-backward asymmetries, we must sum over the 
contributing quarks and fold with the parton  luminosities. \s

A few points are worth recalling concerning the forward-backward asymmetry in
$E_6$ models \cite{AFB}: $(i)$ since the up-type quarks have no axial couplings
to the $Z'$ boson, $Q_3^q=0$, they do not contribute to $A_{\rm FB}^\ell$ on 
the $Z'$ peak; $(ii)$ the asymmetry completely vanishes for three $\beta$
values: $\beta={\rm arctan}(-\sqrt{3/5})$ and $\beta= \pm \pi/2$, where the
left- and right-handed $Z'$ couplings of both $d$-quarks and charged leptons
are equal; $(iii)$ off the $Z'$ resonance,  there is always an asymmetry that
is generated by the $Z$ boson couplings.

\subsection{Observables sensitive to $Z'$ properties}

The LHC discovery potential for a $Z'$ as a mass peak above a small background
in the  reaction $pp\rightarrow Z' \rightarrow \ell^+\ell^-$, with $\ell=e,\mu$,
is well known. The required luminosity to discover  a $Z'$ basically depends
only on its cross section, and therefore on  its mass and couplings. Experimental
effects due to mass resolution, assuming the design parameters of ATLAS or CMS
\cite{Exp1,Exp2}, are known to result in an only  minor reduction of  the
sensitivity.\s

Once a $Z'$ boson is observed at the LHC, we will obviously measure its mass,
its total width and cross section. Furthermore, forward-backward charge
asymmetries on and off the $Z'$ resonance provide additional information  about
its couplings and interference effects with the $Z$ boson and  the photon.  
In addition one can include the analysis of the $Z'$ rapidity distribution, 
which is sensitive to the $Z'$ couplings to $u\bar{u}$ and $d\bar{d}$ quarks.
Such future measurements can be performed as follows at the LHC: \bigskip

\textit{The total decay width of the $Z'$}  is obtained from a fit to the
invariant mass distribution of the reconstructed dilepton system using a
non-relativistic Breit-Wigner function: $a_0/[(M_{\ell \ell}^2-M_{Z'}^2)^2+
a_1]$ with  $a_1 = \Gamma_{Z'}^2 M_{Z'}^2$. \bigskip

\textit{The $Z'$ cross section times leptonic branching ratio}  is calculated
from the number of reconstructed dilepton events lying within $\pm 3\Gamma$
around the observed peak. The $3\Gamma$ interval  used to define the cross
section is arbitrary; however, if varied from $2\Gamma$ to $5\Gamma$, the
cross section  increases only between 5 and 10\% for different $Z'$ models
and masses\footnote{As noted previously, both the total width and the cross
section times the leptonic branching ratio can be altered if exotic decays of
the  $Z'$ boson are present. However, this dependence disappears in the
product, and it is this quantity that should be used in discriminating models
independently of the  decays.}. \s

\begin{figure}[htbp]
\begin{center}
\vspace*{-5mm}
\mbox{
\includegraphics[width=0.52\textwidth,height=7cm]{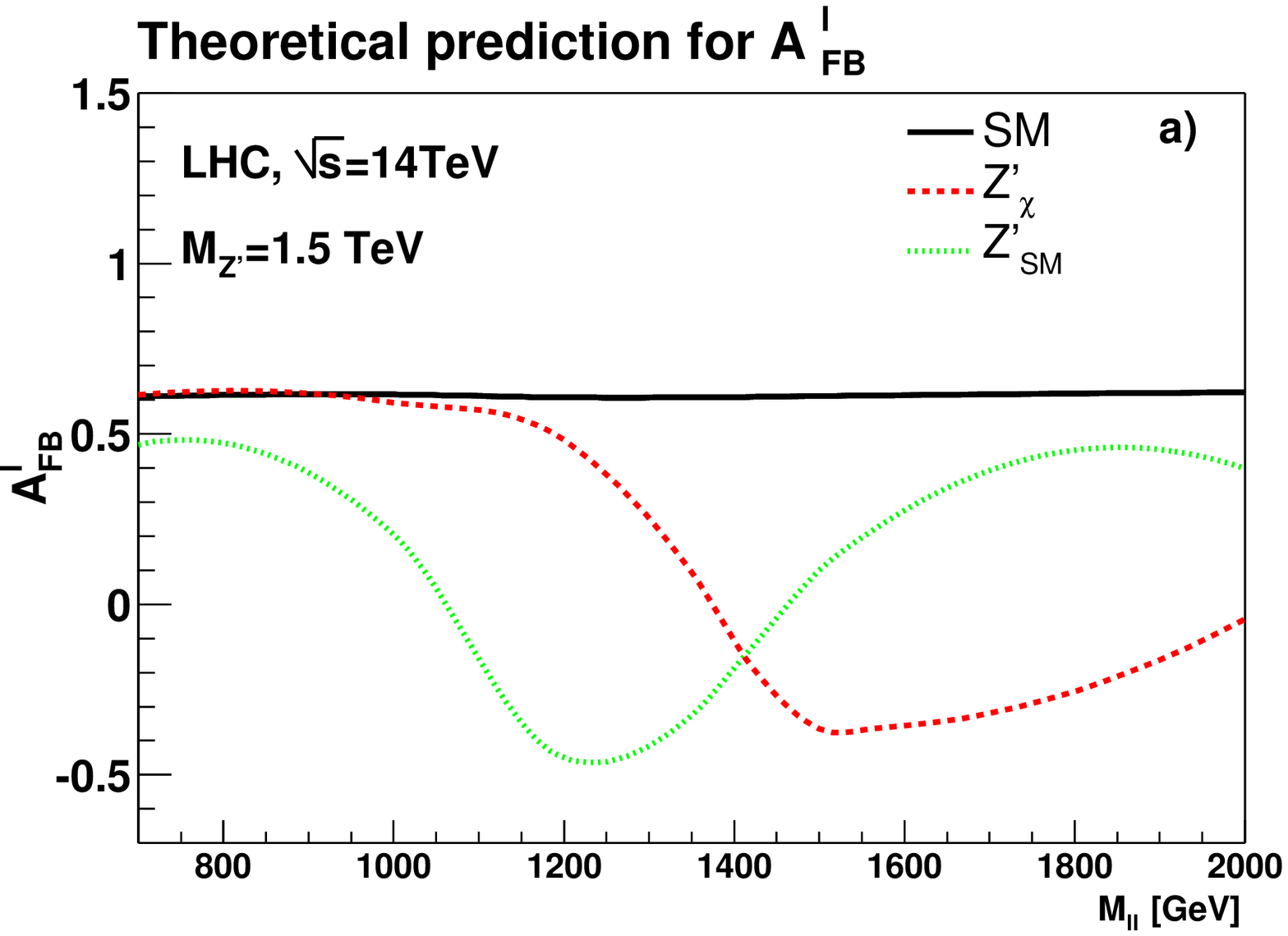}
\includegraphics[width=0.52\textwidth,height=7cm]{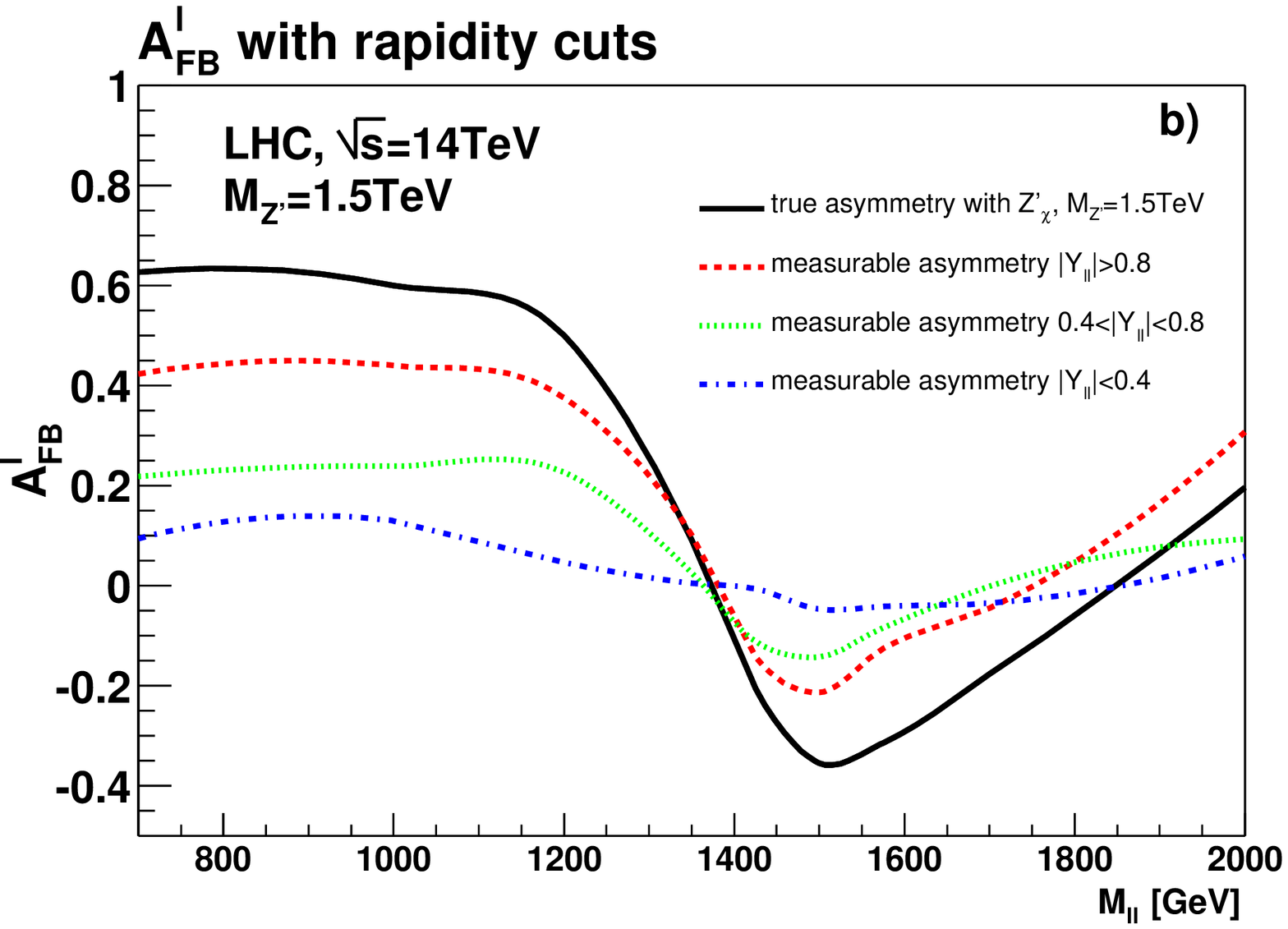}
}
\end{center}
\vspace*{-7mm}
\caption{\it $A_{FB}^\ell$ as a  function of the dilepton mass: the predicted
asymmetry for a $Z'_\chi$ and a $Z'_{SM}$ together with the SM prediction 
(a) and in different dilepton intervals (b).}
\label{afb_mass}
\vspace*{1mm}
\end{figure}

\textit{The leptonic forward-backward charge asymmetry} 
$A_{\rm FB}^{\ell}$ is defined from the lepton angular distribution with respect to the quark 
direction in the centre-of-mass frame, as:
\begin{equation}
\mathrm{\frac{d\sigma}{d \cos \theta ^*} \propto \frac{3}{8} (1 + \cos^2 \theta ^*)
+ A_{\rm FB}^{\ell} \cos \theta ^*} \ .
\end{equation}
As discussed in Ref.~\cite{Previous},  the lepton angle  $\theta^*$ in the
dilepton c.m.\, frame can be calculated  using the measured four momenta of the
dilepton system. $A_{\rm FB}^{\ell}$ can then be determined with an unbinned
maximum likelihood fit to the $\cos \theta ^*$ distribution.  Figure 2a shows
the asymmetry, assuming that the quark direction is known, as a function of the
dilepton mass for a $Z_\chi'$ and a $Z_{\rm SM}'$ boson, assuming a mass
$M_{Z'}=1.5$ TeV. $A_{\rm FB}^{\ell}$ varies strongly with the dilepton mass
and is very  different in the two models. Unfortunately, $A_{\rm FB}^{\ell}$
cannot be measured directly in a proton-proton collider, as  the original quark
direction is not known. However, it can be extracted from  the kinematics of
the dilepton system, as was shown in detail in \cite{Previous}. The method is
based on  the different $x$ spectra of the quarks and antiquarks in the proton,
which allows to approximate the quark direction with  the boost direction of
the $\ell \ell$ system with respect to the beam axis (the $z$ axis).
Consequently, the probability to assign the correct quark direction increases
for larger rapidities of the  dilepton system and somewhat  cleaner and more
significant measurements can be performed, as shown in Fig.\,\ref{afb_mass}b. 
A purer, though smaller, signal sample can thus be obtained by introducing a
rapidity cut. For the following studies we will require $|Y_{\ell\ell}|>0.8$. 

\textit{The $Z'$ rapidity distribution} allows us to   obtain the fraction of
$Z'$ bosons produced from  $u\bar{u}$ and $d\bar{d}$ initial states. Assuming that the
$W^{\pm}$ and $Z$ boson rapidity distributions have been measured  in detail,
as discussed in \cite{lhclumi}, relative parton distribution functions for
$u$ and $d$ quarks, as well as for the corresponding  sea quarks and
antiquarks are well known. Thus, the rapidity spectra can be calculated
separately for  $u\bar{u}$ and $d\bar{d}$, as well as for  sea quark
antiquark annihilation, and for the mass region of interest. Using these
distributions, a fit can be performed to the $Z'$ rapidity distribution,
which allows to obtain the corresponding fractions of the $Z'$  boson
produced from $u\bar{u}$, $d\bar{d}$ as well as for  sea quark-antiquark
annihilation\footnote{Following this procedure, and having  very large
statistics at hand, it would be imaginable even to measure also the  
forward-backward charge asymmetries separately for $u$ and $d$ quarks. 
Charge asymmetries for different $Z'$ rapidity intervals would have  to be
measured and, with the knowledge of the corresponding $u\bar{u}$ and
$d\bar{d}$ fractions  from the entire rapidity distribution,  the
corresponding $u$ and $d$ asymmetries could eventually be disentangled.
However, a quick analysis  of the potential sensitivity indicates that an
interesting statistical sensitivity   would require a luminosity of at least
1000 fb$^{-1}$.}.  \bigskip

In the present analysis, {\tt PYTHIA} events of the type $pp \to \gamma,Z,Z'
\to ee,\mu\mu$ were simulated at a centre-of-mass energy of 14~TeV, and for the
$Z'$ models discussed in section \ref{models} The $Z'$ masses were varied from
1~TeV up to 5~TeV. These events were analysed, using simple acceptance cuts
following  the design criteria of ATLAS and CMS. Following the results from
previous studies and the expected excellent detector resolutions, the obtained
values are known to be   rather insensitive to measurement errors, especially
for the $e^{+}e^{-}$ final states.  We therefore do not include any resolution
for the current study.  In detail, the following basic event selection criteria
were used: 

\begin{itemize}
\vspace*{-3mm} 
\item The transverse momenta of the leptons, $p_T^\ell$\,, should be at least 
20~GeV.
\vspace*{-3mm}

\item The pseudorapidity $|\eta|$ of each lepton should be smaller than 2.5.
\vspace*{-3mm}

\item The leptons should be isolated, requiring that the lepton carries at
least 95\% of the total transverse energy  found in a cone of size of 0.5
around the lepton. 
\vspace*{-3mm}

\item There should be exactly two isolated leptons with opposite charge 
in each event.
\vspace*{-3mm}

\item The two leptons should be back to back in the plane transverse to the
beam direction, so that the opening angle between them was larger than
 $160^{\circ}$.
\vspace*{-3mm}
 \end{itemize}

\begin{figure}[htbp]
\vspace*{-5mm}
\begin{center}
\includegraphics[width=0.7\textwidth]{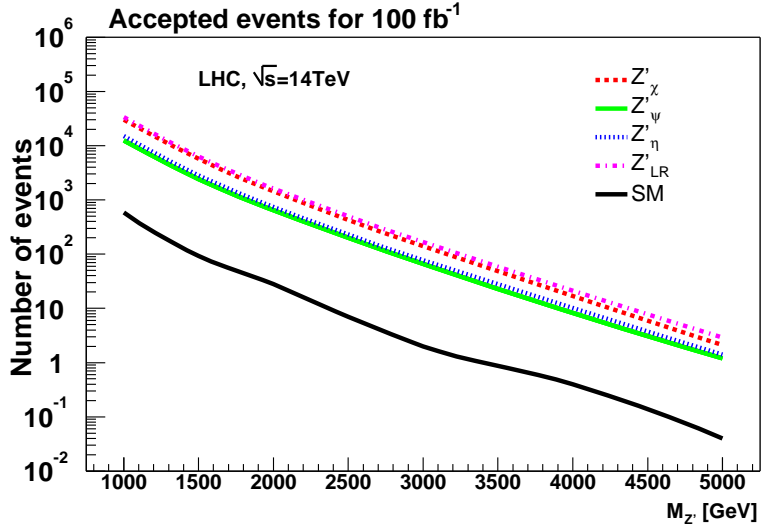}
\end{center}
\vspace*{-1cm}
\caption{\it  The LHC discovery reach: the number of events fulfilling the
selection cuts for a  luminosity of 100~fb$^{-1}$, for the different $Z'$
models and in the SM case.}  \label{disc}
\vspace*{-2mm}
\end{figure}

Figure \ref{disc} shows the expected number of events for a luminosity of
100~fb$^{-1}$. The SM background relative to the signal cross section is
found to  be essentially negligible for the considered $Z'$ models. We thus
reconfirm the known $Z'$ boson LHC discovery potential,  to reach masses up
to about 5~TeV for a luminosity of 100 fb$^{-1}$ \cite{Theory}.\s 

\begin{figure}[htbp]
\vspace*{-4mm}
\begin{center}
\mbox{
\includegraphics[width=0.52\textwidth,height=8cm]{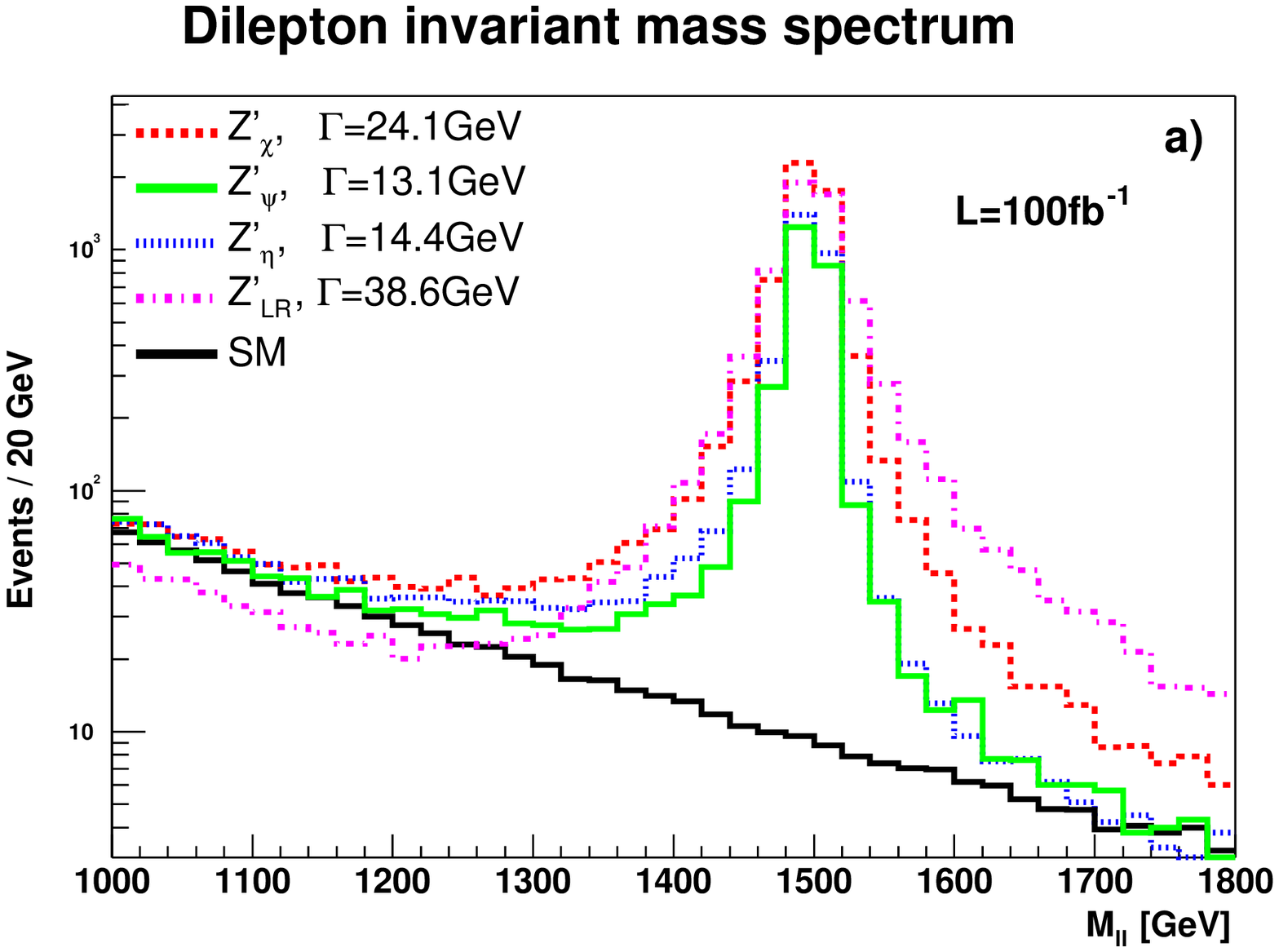}
\includegraphics[width=0.52\textwidth,height=8cm]{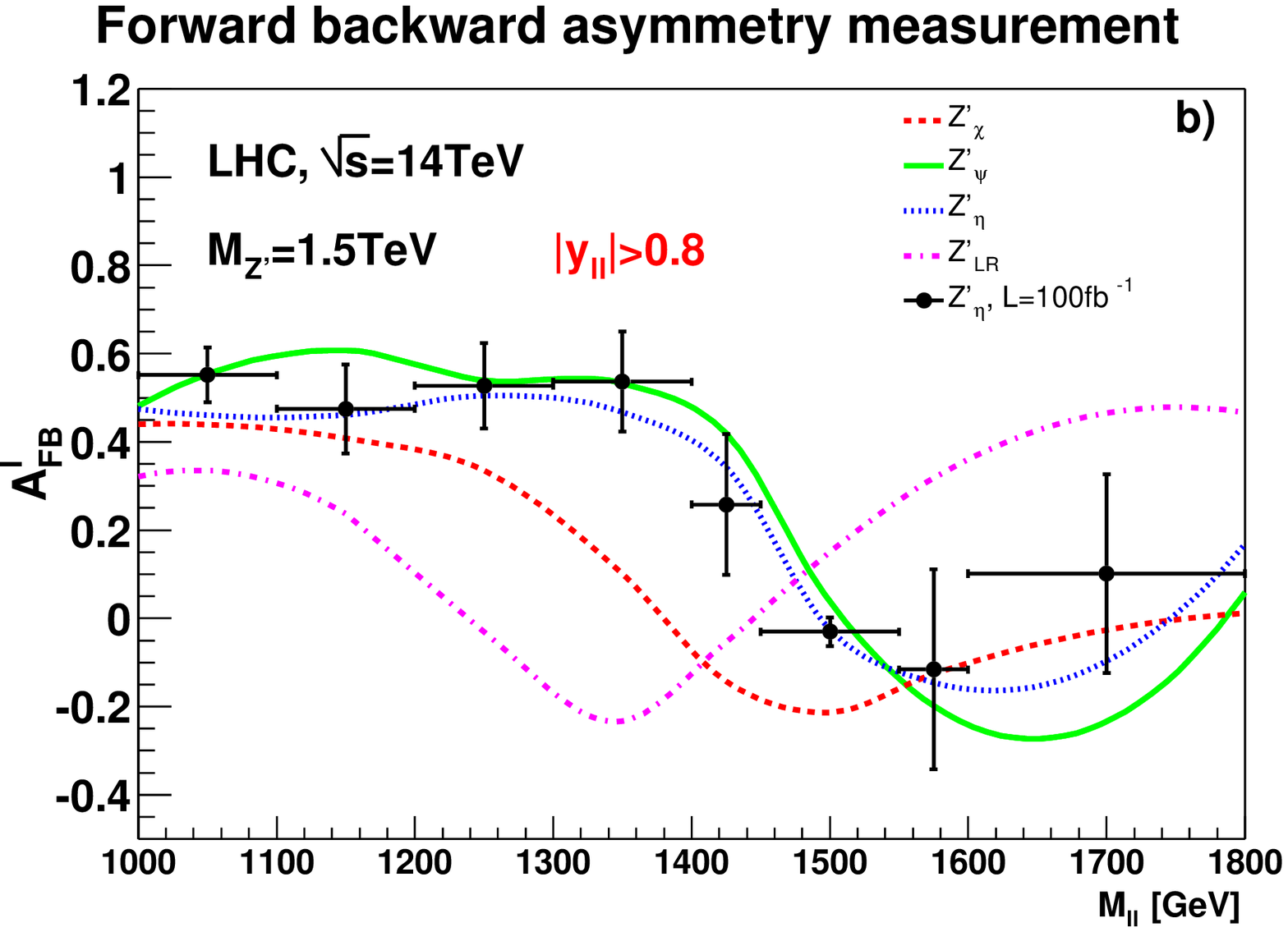}
}
\end{center}
\vspace*{-5mm}
\caption{\it The dilepton invariant mass spectrum (a) and $A^{\ell}_{FB}$ (b)
as a function of $M_{\ell \ell}$ for four $Z'$ models. 
For the forward-backward charge asymmetry, the 
rapidity of the dilepton system is required to be larger than 0.8.
A simulation of the statistical errors, including random fluctuations of the 
$Z'_{\eta}$ model and with errors 
corresponding to a luminosity of 100~fb$^{-1}$ has been included 
in (b). }
\label{fakeexp}
\vspace*{-3mm}
\end{figure}

Figure \ref{fakeexp}a shows the invariant mass distribution for the dilepton
system, as expected for different models with $M_{Z'}$ fixed to 1.5 TeV and for
the SM using a luminosity of 100 fb$^{-1}$. For all $Z'$ models, huge peaks,
corresponding to 3000--6000 signal events,  are found above a small background.
The cross sections for $Z'$ bosons in the various models are also  strongly
varying.\s

In addition, very distinct observable forward-backward charge asymmetries are
expected as a function of the dilepton mass and   for the different $Z'$
models, as shown in Fig.~\ref{fakeexp}b. In order to get an impression of how
an experimental signal with statistical fluctuations would look like, the
measurable forward-backward asymmetry in the $Z'_{\eta}$ case has been
generated  with the number of events corresponding to 100~fb$^{-1}$, as shown
in Fig.~\ref{fakeexp}b.  We find that additional and complementary
informations is also obtained  from $A_{\rm FB}^{\ell}$ measured in the
interference region. To quantify the study for a $Z'$ mass of 1.5~TeV, 
``on-peak events" are counted if the dilepton mass is found in the interval
$1.45$~TeV\ $\leq M_{\ell \ell} \leq$ 1.55~TeV.  The ``interference region" is
defined  accordingly and satisfy 1~TeV\, $\leq M_{\ell \ell} \leq$~1.45~TeV.\s

\begin{figure}[htbp]

\vspace*{-2mm}
\begin{center}
\mbox{
\includegraphics[width=0.52\textwidth,height=8cm]{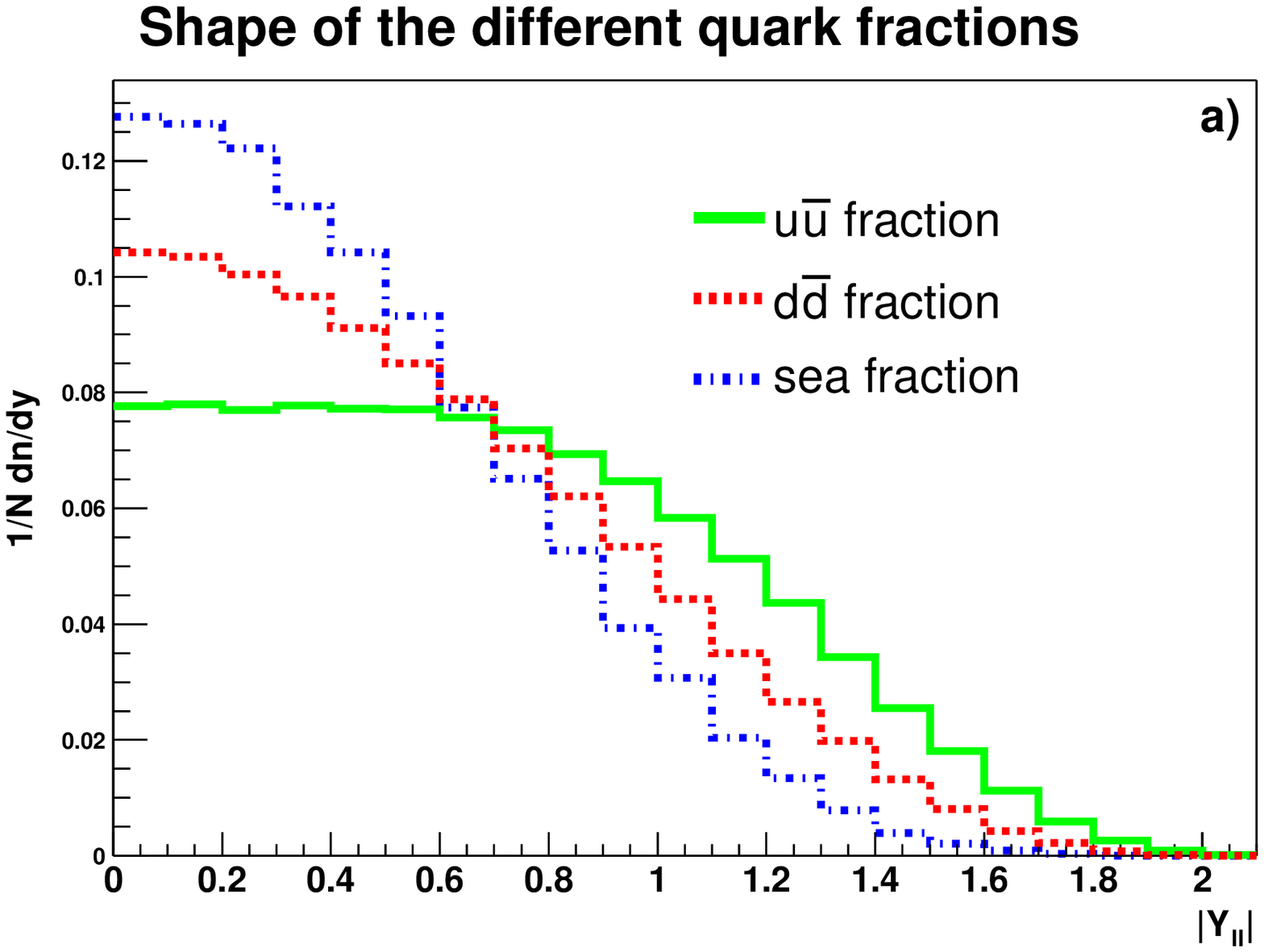}
\includegraphics[width=0.52\textwidth,height=8cm]{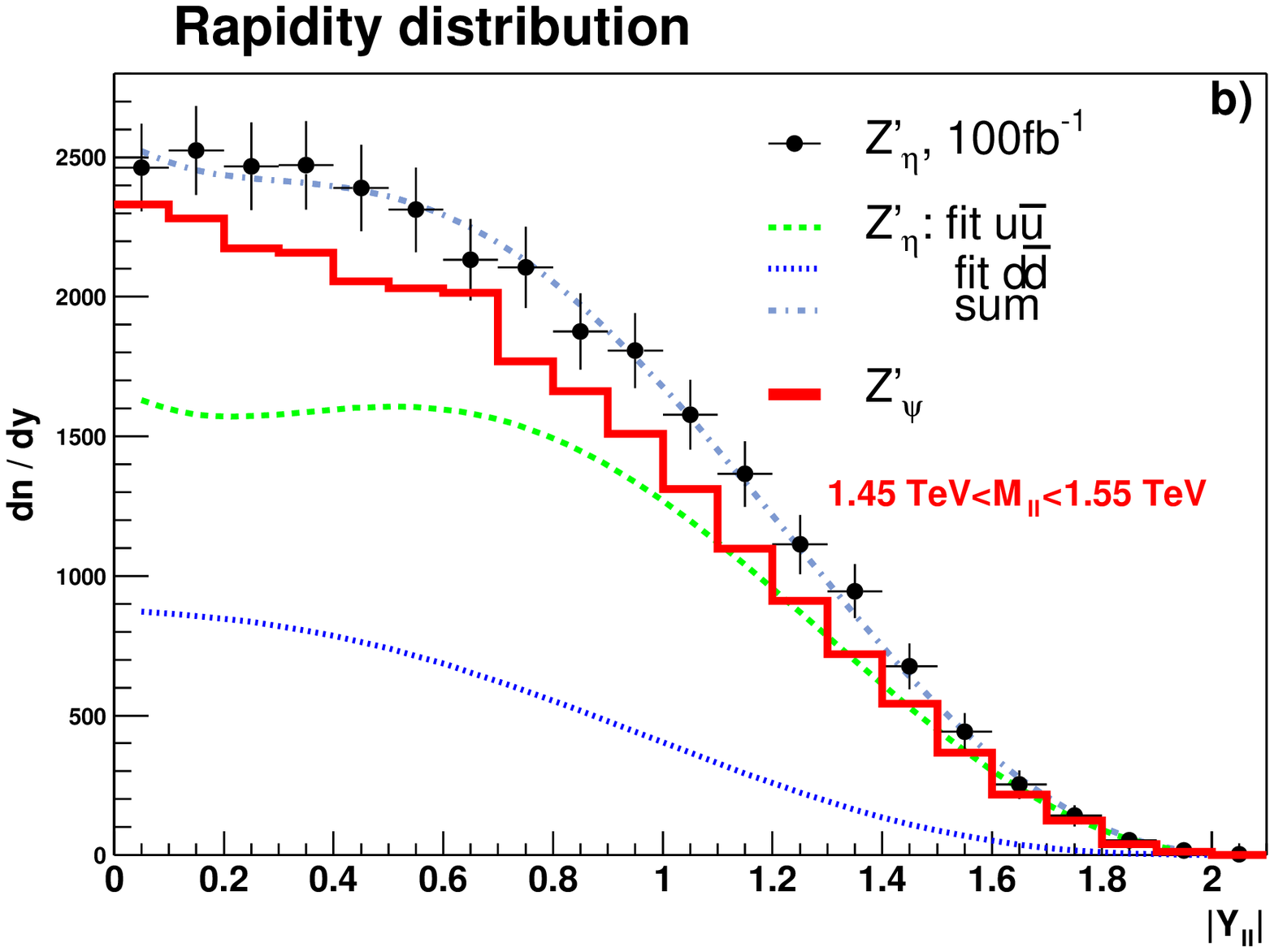}
}
\end{center}
\vspace*{-5mm}
\caption{\it The normalized rapidity distribution of $Z'$ with a mass of $1.5
\pm 0.05$ TeV produced from the different types of quarks (a). The observable
rapidity distribution for two different $Z'$ models is shown in (b),  including
the fit results that determine the different types of $q\bar{q}$ fractions.} 
\label{rap}
\vspace*{-3mm}
\end{figure}

Finally, the rapidity distribution is analysed.  Figure \ref{rap}a shows the
normalized distributions for  a $Z'$ with a mass of 1.5 TeV produced from
$u\bar{u}$, $d\bar{d}$ and  sea-antisea quark annihilation. Especially the $Z'$
rapidity distribution from  $u\bar{u}$ annihilation appears to be significantly
different from the other two distributions.  Figure \ref{rap}b shows the
expected rapidity distribution  for the $Z'_{\eta}$ model. A particular $Z'$
rapidity distribution is fitted using a linear combination of the three pure
quark-antiquark rapidity distributions  shown in Fig.~\ref{rap}b. The fit
output gives the  $u\bar{u}$, $d\bar{d}$ and $sea$ quarks fraction in the
sample. This will thus reveal how the $Z'$ couples to different quark flavours
in a particular model. \s

In order to demonstrate the analysis power of this method,  we also show the
rapidity distribution in the case of the $Z'_{\psi}$ boson, which has equal
couplings to up-type and down-type quarks.  As can be qualitatively expected
from the distributions shown in Fig.~5a,  the used fitting procedure provides
very accurate results for the known generated  fraction $R_{u\bar{u}}$ of
$u\bar{u}/$all, while some correlations between $d\bar{d}$ and the
sea-antisea  $Z'$ production, which limits the accuracy of the measurement
for the $d\bar{d}$ fractions. For example, for the $Z'_{\eta}$ model, the
generated event fractions from  $u\bar{u}$, $d\bar{d}$ and sea-antisea quarks
are 0.71, 0.26 and 0.03 respectively. The corresponding numbers from the fit
and 100 fb$^{-1}$ are 0.71$\pm$0.07, 0.29$\pm$0.08 and 0.01$\pm$0.02.\s

Table~\ref{moddisc} shows the value of the cross section times the total decay
width, the forward-backward charge asymmetry for the on-peak and interference
regions as defined  above, and the ratio of $Z'$ events produced from $u\bar{u}$
annihilation  as obtained from the fit to the $Z'$ rapidity distribution.

\begin{table}[htbp]
\renewcommand{\arraystretch}{1.25}
\begin{center}
\begin{tabular}{|c||rcl||rcl|rcl||rcl|}
\hline \hline
Model &  \multicolumn{3}{c||}{$\sigma^{3\Gamma}_{ll} \times \Gamma$
[fb$\cdot$GeV]} 
        & \multicolumn{3}{c|}{$A^{\rm on-peak}_{\rm FB}$} 
        & \multicolumn{3}{c||}{$A^{\rm off-peak}_{\rm FB}$}
        & \multicolumn{3}{c|}{$R_{u\bar{u}}$} \\
\hline 
\hline
$Z'_{\psi}$ & 487 &$\pm$& 5 & 0.04 &$\pm$& 0.03 & 0.53 &$\pm$& 0.04 & 0.60 &$\pm$& 0.07 \\
\hline
$Z'_{\eta}$  & 630 &$\pm$& 20 & $-0.03$ &$\pm$& 0.03 & 0.45 &$\pm$& 0.04 & 0.71 &$\pm$& 0.07 \\
\hline
$Z'_{\chi}$  & 2050 &$\pm$& 40 & $-0.23$ &$\pm$& 0.02 & 0.26 &$\pm$& 0.05 & 0.22 &$\pm$& 0.05 \\
\hline
$Z'_{LR}$  & 3630 &$\pm$& 80 & 0.15 &$\pm$& 0.02 & 0.06 &$\pm$& 0.06 & 0.45 &$\pm$& 0.05 \\
\hline 
$Z'_{SM}$  & 8000 &$\pm$& 140 & 0.07 &$\pm$& 0.02 & 0.18 &
$\pm$& 0.03 & 0.05 &$\pm$& 0.04 \\
\hline
$Z'_{d}$  & 1520 &$\pm$& 40 & $-0.50$ &$\pm$& 0.02 & 0.26 &
$\pm$& 0.05 & 0.00 &$\pm$& 0.01 \\
\hline
\hline
\end{tabular} 
\end{center}
\vspace*{-3mm}
\caption{\it The values of the four basic observables, the signal cross
section, multiplied  by the total width, the forward-backward charge asymmetry
on- and off-peak, and the ratio $R_{u\bar{u}}$ for various $Z'$ models and with
a $Z'$ mass of 1.5 TeV. The quoted statistical errors are those that can be
expected for a luminosity of 100~fb$^{-1}$.}  \label{moddisc}

\end{table}

\subsection{Distinction between models and parameter determination}

Let us now discuss how well the different $Z'$ models can be distinguished
experimentally   using the observables  defined before: $\sigma^{3 \Gamma}
_{\ell\ell} \times \Gamma$, $A_{\rm FB}^{\ell}$ on- and off-peak, as well as
$R_{u\bar{u}}$ as obtained from the rapidity distribution.  As a working
hypothesis, a luminosity of 100~fb$^{-1}$ and a $Z'$ mass of 1.5~TeV  will be
assumed in the following. \s

A precise knowledge of the cross section times the total width allows a first
good distinction to be made between some models, as shown in the upper two
plots of  Fig.~6. It is not obvious how accurately absolute cross sections
can be measured and interpreted at the LHC. However, following the procedure
outlined in \cite{lhclumi}, comparable reactions, in this  case  $Z'$ and $Z$
boson production, should be counted with respect to each other.  The use of
such ratio measurements should allow us to minimize systematic uncertainties,
and an accuracy of $\pm$1\% might be achievable. As can be seen from the
other plots in Fig.~6, the additional variables show a different sensitivity
for the different couplings. \s

For example, very similar cross sections are expected for the $E_6$ $Z'$
models with  $\cos \beta \sim \pm 1$ and for left-right models with
$\alpha_{LR} \lsim  1.3$. However, these two different models show a very
different  behaviour for on- and especially off-peak asymmetries and for the
couplings to up-type and down-type quarks. Obviously, the maximum sensitivity
can be obtained by using all observables together.  Having said this, one
also needs to point out that some ambiguities between the  different models
remain, even after a complete analysis of 100 fb$^{-1}$ of data.\s

Assuming that a particular model has been selected, one would like to know 
how well the parameter(s), such as $\cos\beta$ or $\alpha_{LR}$, can be
constrained.  In the case of the $E_6$ model for instance, one finds that
$\cos\beta$ cannot always be determined unambiguously. Very similar results
can be expected for different observables  but using very different values
for $\cos\beta$. Again, the combination of the various measurements helps to
reduce some ambiguities.

\begin{figure}[htbp]
\begin{center}
\vspace*{-7mm}
\mbox{
\includegraphics[width=0.5\textwidth, height=5.5cm]{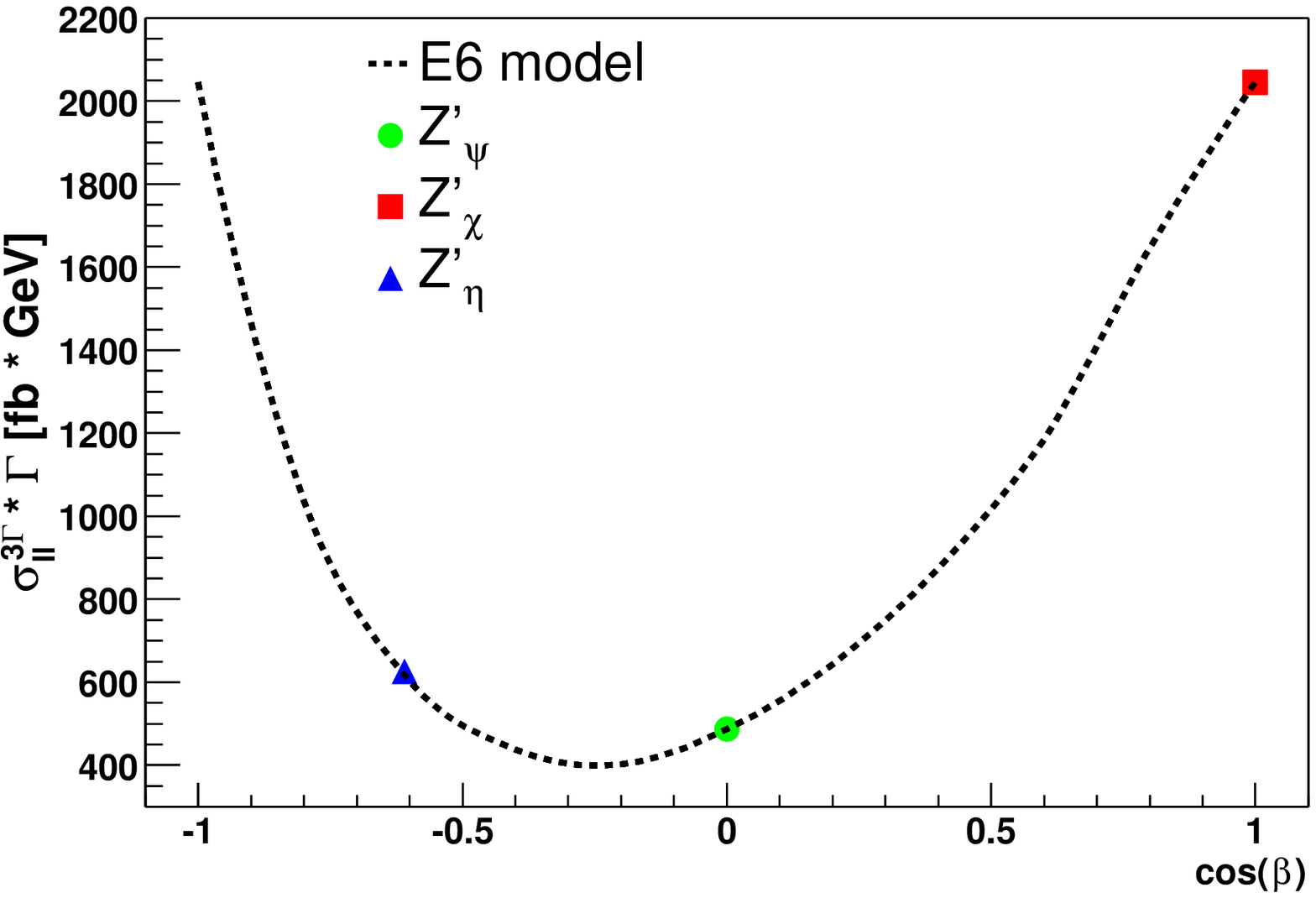}
\includegraphics[width=0.5\textwidth, height=5.5cm]{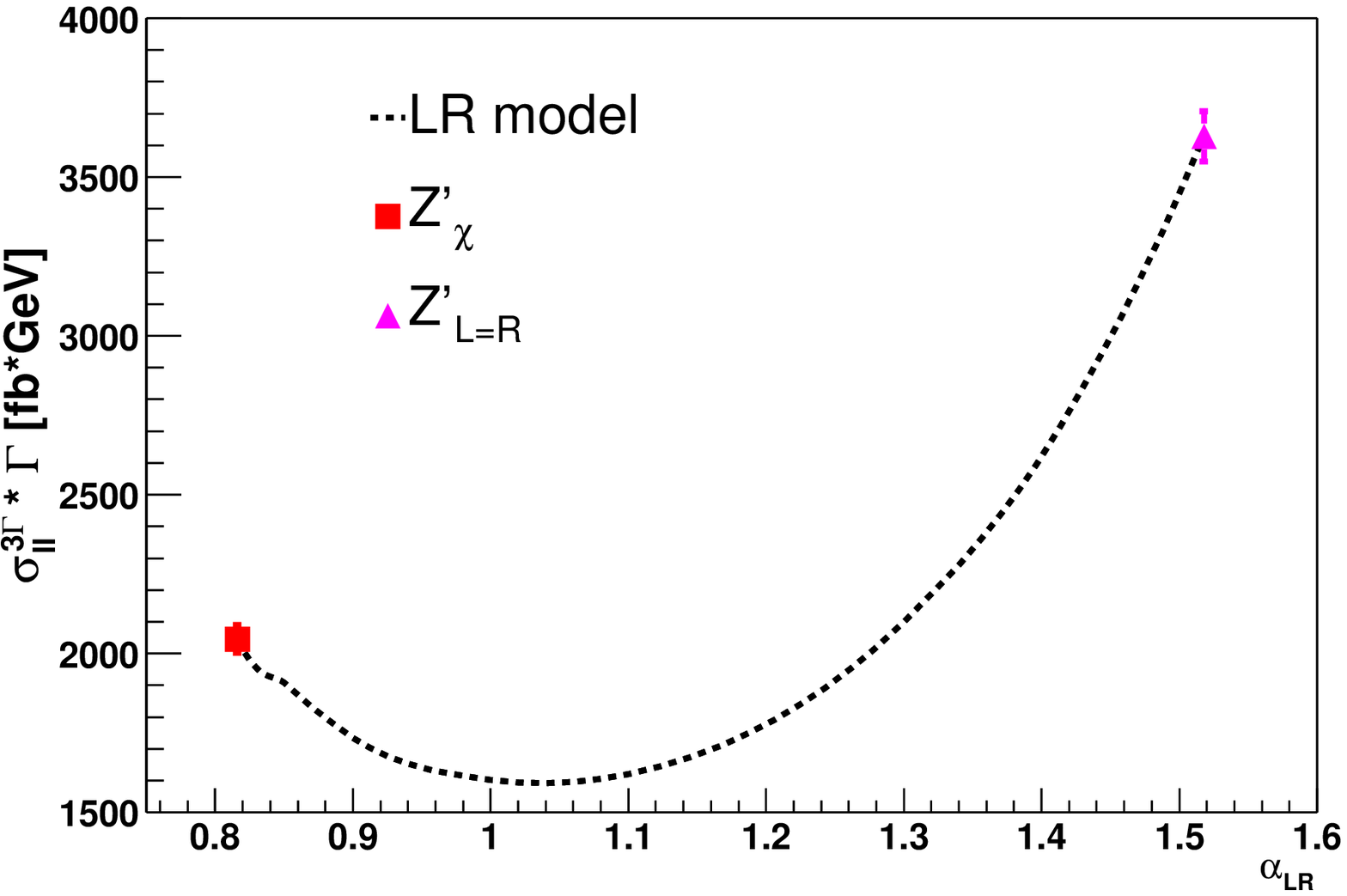}
}
\vspace*{-3mm}
\mbox{
\includegraphics[width=0.5\textwidth, height=5.5cm]{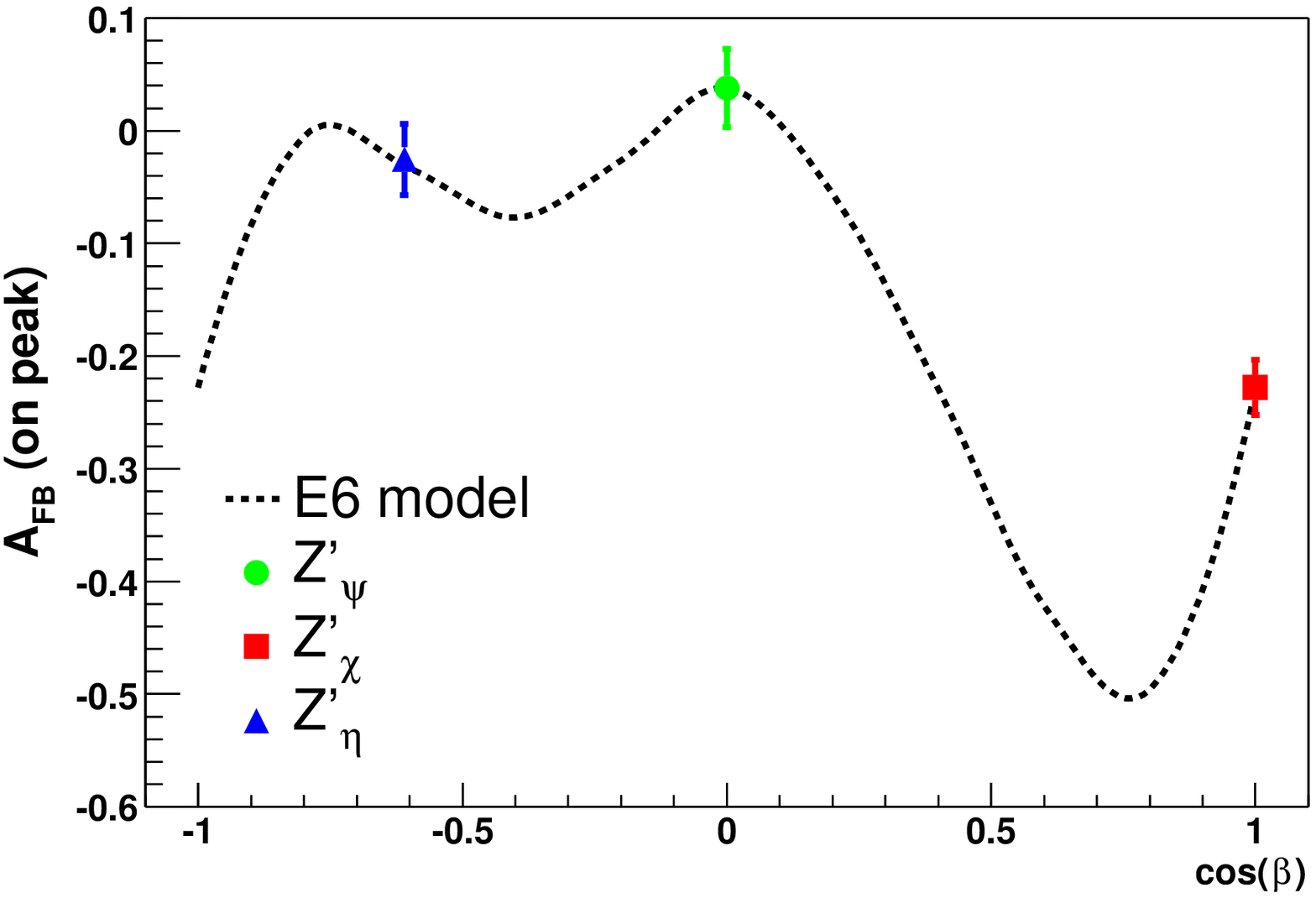}
\includegraphics[width=0.5\textwidth, height=5.5cm]{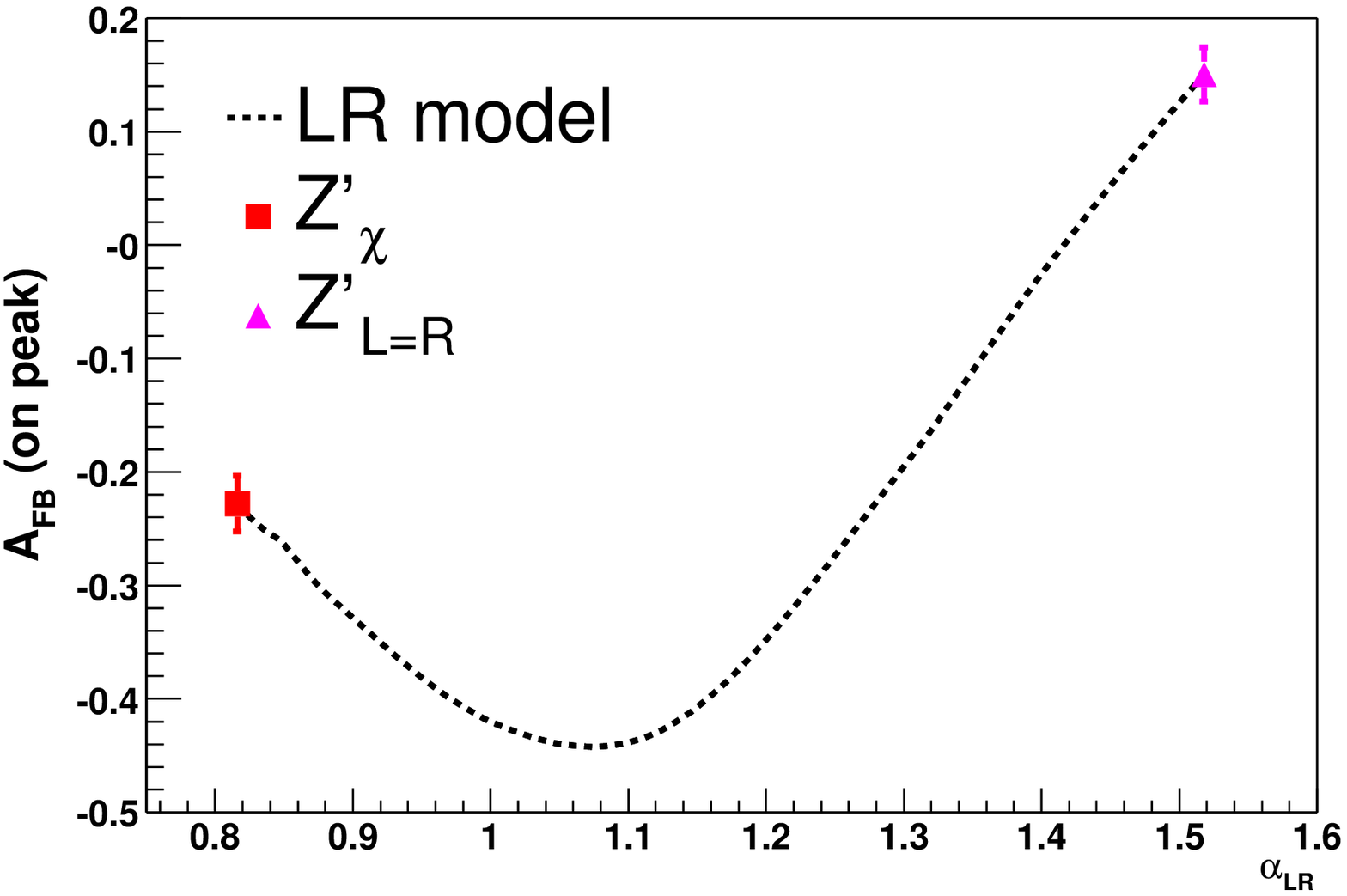}
}
\vspace*{-3mm}
\mbox{
\includegraphics[width=0.5\textwidth, height=5.5cm]{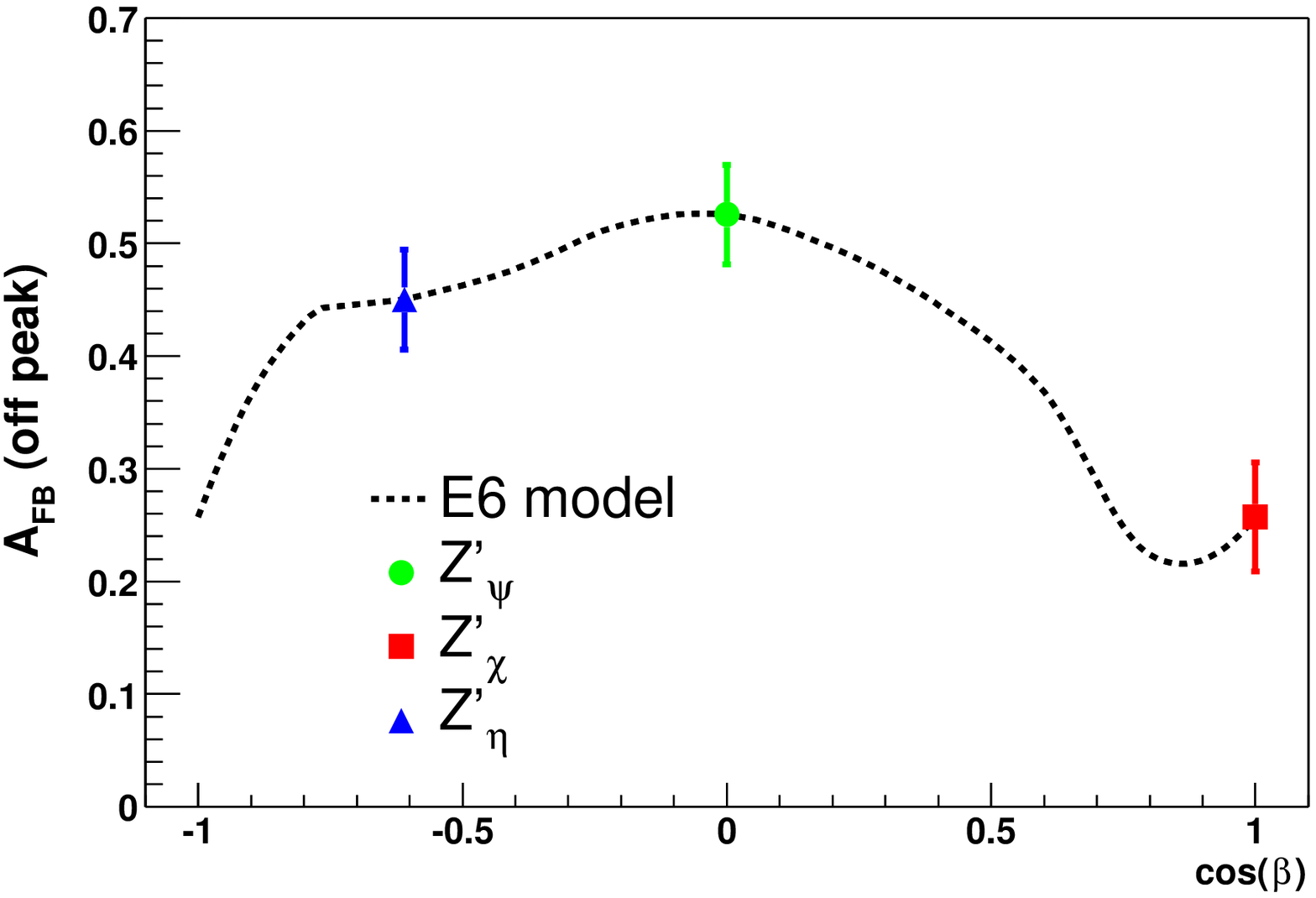}
\includegraphics[width=0.5\textwidth, height=5.5cm]{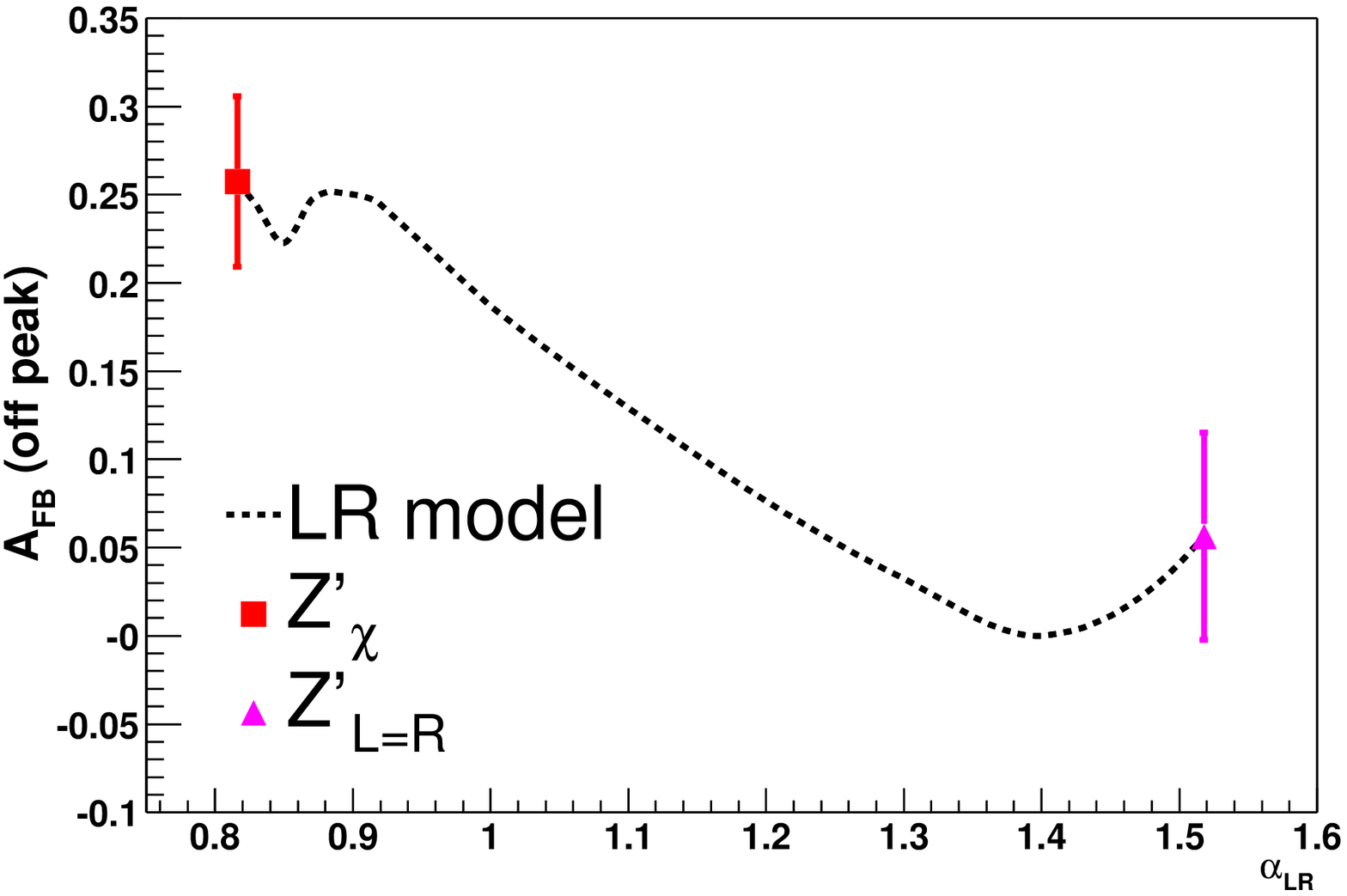}
}
\vspace*{-3mm}
\mbox{
\includegraphics[width=0.5\textwidth, height=5.5cm]{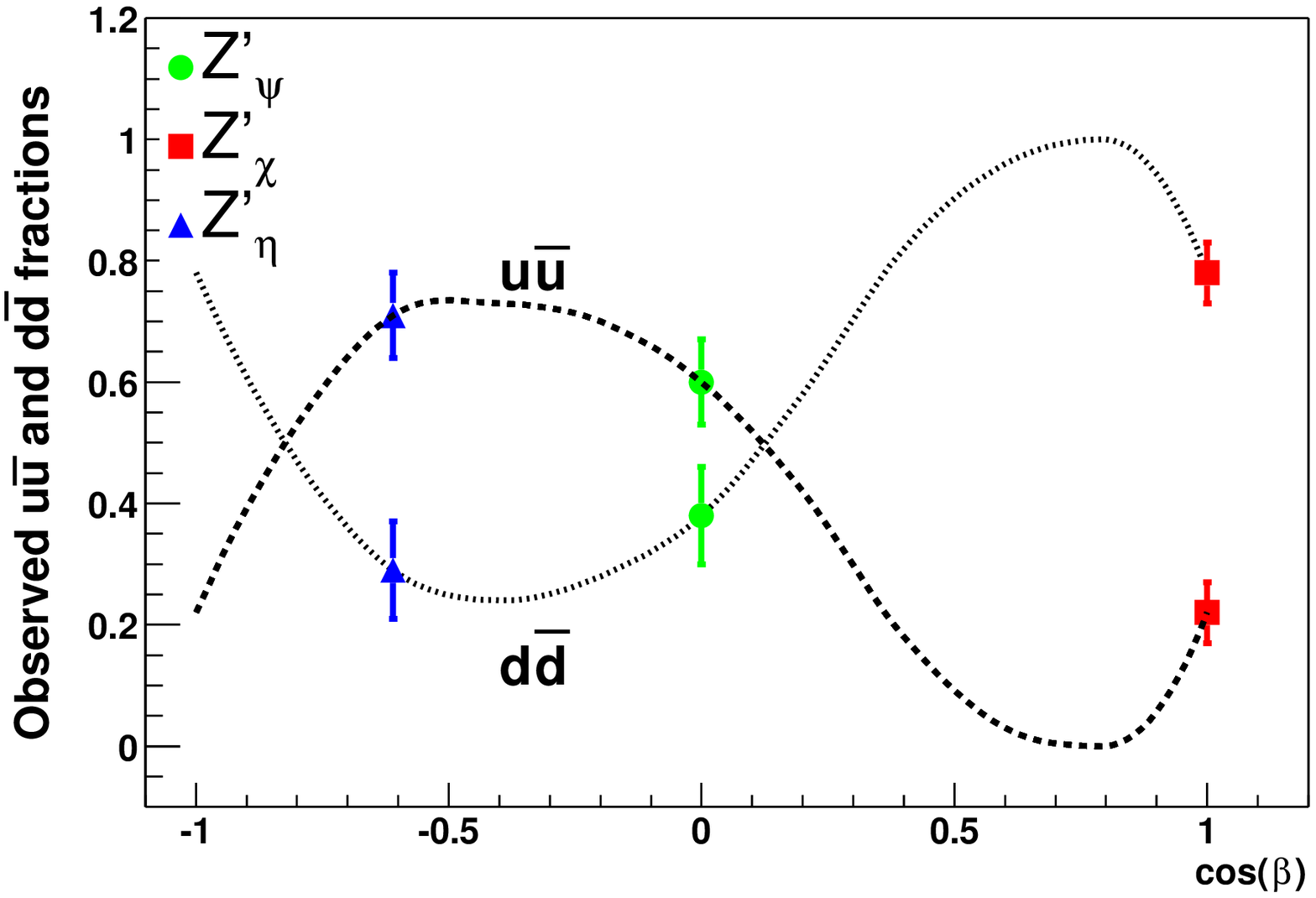}
\includegraphics[width=0.5\textwidth, height=5.5cm]{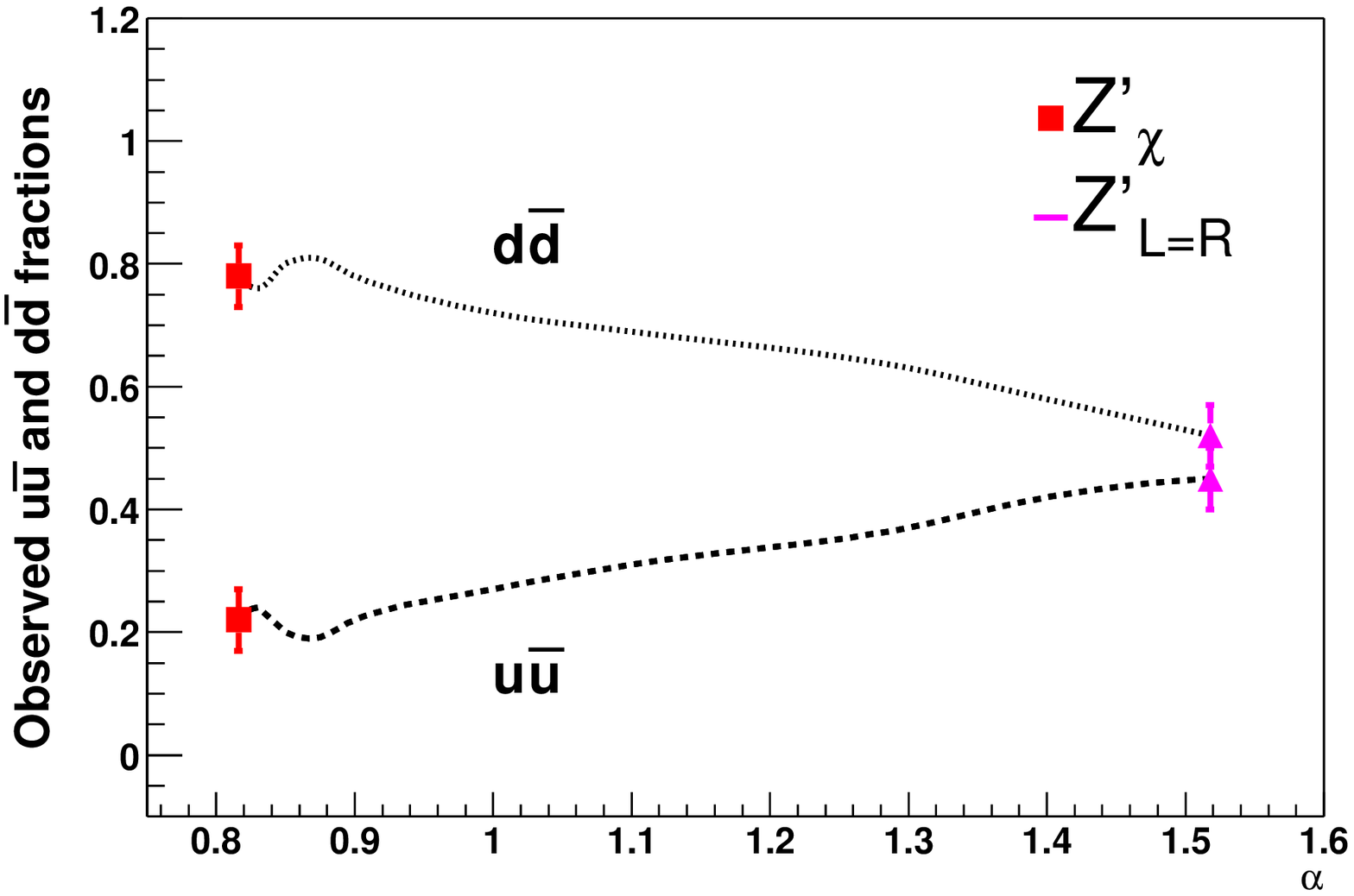}
}
\end{center}
\vspace*{-3mm}
\caption{\it Variation of $\sigma^{3\Gamma}_{\ell\ell} \cdot \Gamma$,  $A^{\rm
on-peak}_{FB}$, $A^{\rm off-peak}_{FB}$ and the ratio $R_{q\bar{q}}$ as a
function of the $E_6$ model parameter $\cos\beta$ (left) and the LR-model
parameter $\alpha_{LR}$. The points corresponding to the particular $Z'$ models
are also shown. }
\label{aroundcosbeta}
\end{figure}

If the $Z'$ mass is increased, the number of events decreases drastically and
the differences between the models start to become covered within the
statistical fluctuations. For the assumed luminosity of 100~fb$^{-1}$, we
could still distinguish a $Z'_{\chi}$ from a $Z'_{LR}$ over a large parameter
range; the  $A_{\rm FB}^{\ell}$ measurements provide some statistical
significance up to $M_{Z'}=2$--$2.5$~TeV. On the contrary, a $Z'_{\eta}$
could be differentiated from a $Z'_{\psi}$ only up to a $Z'$ mass of at most
2~TeV as, in that case, the dependence of $A_{\rm FB}^{\ell}$  is almost
identical in the two models.\bigskip

In summary, a realistic simulation of the study of the properties of  $Z'$
bosons originating from various theoretical models has been performed for the
LHC. We have shown that, in addition  to the $Z'$ production cross section
times total decay width,  the measurement of the forward-backward lepton charge
asymmetry, both on the $Z'$ peak and in the interference region, provide
complementary information. We have also shown that a fit of the rapidity
distribution can provide a sensitivity to the $Z'$ couplings to up-type and
down-type quarks. The combination of all these observables would allow us to
discriminate between $Z'$ bosons of different models or classes of models for
masses up to 2--2.5 TeV, if a luminosity of 100 fb$^{-1}$ is collected.

\bigskip 

\nn {\bf Acknowledgements:} We thank J. Hewett, G. Polesello and T. Rizzo for
helpfull discussions during the Les Houches Workshop.

\end{document}